\documentclass[conference]{IEEEtran}

\usepackage{balance}
\usepackage{comment}
\usepackage{subfigure}
\usepackage{graphicx}
\usepackage{epsfig}
\usepackage{url}
\usepackage{amsfonts}
\usepackage{amssymb}
\usepackage{mdwmath}
\usepackage{mdwtab}
\usepackage{cite}
\usepackage{booktabs}
\usepackage{tabularx}
\usepackage{multirow} 
\usepackage[]{caption}
\usepackage{amsmath}
\usepackage{graphicx}
\usepackage{hyperref}
\usepackage{xcolor}
\usepackage{relsize}

\begin{document}
\title{An Interoperable Realization of Smart Cities with Plug and Play based Device Management}
\author{
\IEEEauthorblockN{Prasant Misra$^\dagger$, Vasanth Rajaraman$^\dagger$, Kumaresh Dhotrad$^\dagger$, Jay Warrior$^\dagger$, Yogesh Simmhan$^{\ddagger}$}
\IEEEauthorblockA{$^{\dagger}$Robert Bosch Centre for Cyber Physical Systems, Indian Institute of Science, Bangalore, India\\
								$^{\ddagger}$Supercomputing Education and Research Centre, Indian Institute of Science, Bangalore, India\\
Email: \{prasant.misra, vasanth.rajaraman, jay.warrior\}@rbccps.org$\dagger$ simmhan@serc.iisc.in$^{\ddagger}$}
}

\maketitle
\begin{abstract}
The primal problem with Internet of Things (IoT) solutions for smart cities is the lack of interoperability at various levels, and more predominately at the device level.
While there exist multitude of platforms from multiple manufacturers, the existing ecosystem still remains highly closed.
In this paper, we propose SNaaS or Sensor/Network as a Service: a service layer that enables the creation of the plug-n-play infrastructure, across platforms from multiple vendors, necessary for interoperability and successful deployment of large-scale city wide systems.
In order to correctly position the new service layer, we present a high level reference IoT architecture for smart city implementations, and follow it up with the workflow details of SNaaS along with preliminary microbenchmarks.
\end{abstract}


\section{Introduction}

The world is undergoing an unprecedented pace of urbanization, and if present trends continue, the world urban population will rise by about $72$\% in the next $4$ decades.
While the reasons for this rapid urban expansion are many, it is primary driven by two entities with complementary needs: \emph{cities}, on one end, to attract the best skilled people and enterprises; and \emph{people}, on the other end, by their preference to migrate to cities that provide better quality of life. 
This rapid scale of urbanization will need smarter, sustainable cities that are able to effectively and efficiently manage city utilities and services for its citizens.
\newline
\indent
Electric grids, water distribution systems, transportation systems, communication infrastructure, waste treatment plants, commercial buildings, hospitals, homes and education centers are existing \emph{vital} facilities that shape the liveability standard of a city. 
In the future, newer infrastructures and services will also bring benefits and create opportunities of added value to its citizens.
An efficient and effective management of these existing and possibly new city systems requires \emph{integration}, an aspect needed for \emph{transforming} a traditional city into a smarter form. 
However, value creation through integration can \emph{only} be delivered with compatibility of technologies that is best achieved through standards that ensure \emph{interoperability}.
\newline
\indent
The technology revolution driven by the Internet of Things, or the IoT offers promising smart city solutions.
IoT, in its basic definition, is a vision for a ubiquitous society wherein people and ``Things'' are connected in an immersively networked computing environment.
However, current IoT-based smart city implementations have mainly concentrated on vertical integration of independent infrastructure and services for siloed applications (Table~\ref{tab:smartcity}).
Available solutions are extremely \emph{closed} with an ecosystem that is highly \emph{locked-in} by vendors, wherein a single vendor owns the vertical application, platform, communication, services, and data.
While interoperability will ensure better system management and open markets to competitive solutions that are affordable and sustainable, the existing ecosystem allows minimal/no flexibility.
\newline
\indent
In this paper, we explore the interoperability problem at the ``Things'' (i.e., device) level.
``Things'' or physical modules/platforms contain the interface to the physical world for measurement of various physical parameters.
Due to interoperability constraints, the existing IoT ecosystem does not have provisions for interfacing devices of varying heterogeneity and complexity from multiple manufacturers that use different proprietary middleware technologies.
In order to overcome this fundamental limitation, we propose SNaaS. 
\newline
\indent
Expanded as Sensor/Network as a Service, SNaaS enables the creation of the plug-n-play infrastructure across platforms from multiple vendors, necessary for interoperability and successful deployment of large-scale city wide systems.
SNaaS abstracts the features, capabilities and controls of the sensors and communication layer to allow higher level services to be developed and interacts with sensors/networks in a generalizable manner. 
In our instantiation of SNaaS, we leverage the IEEE~$1451$\cite{1451} standard (originally developed for industrial processes automation) for describing the capabilities of the sensors/networks.
\newline
\indent
The rest of the paper unfolds as follows.
To ground our discussion, we provide a high level reference IoT architecture to understand the rightful placeholder of SNaaS in the smart city framework in the next section. 
It is followed by Section~\ref{sec:snaas} that presents the architecture and implementation details of SNaaS with preliminary microbenchmarks.

\begin{table*}[t]
  \begin{center}
  \caption{\textbf{IoT based solutions for siloed smart city applications}}
	\begin{tabular}[c]{p{4cm}ccccccccc}
	\toprule
	& {\bf Smart} & {\bf Smart} & {\bf Traffic} & {\bf Smart} & {\bf Pollution} & {\bf Smart Street} & {\bf Smart} & \bf {Smart} & {\bf Smart}\\
 &  {\bf Bus Stop} & {\bf Parking} & {\bf Management} & {\bf Grid} & {\bf Monitoring} & {\bf Lightning} & {\bf Garbage} & \bf{Healthcare} & {\bf Citizen}\\
\midrule
\bf Barcelona, Spain & \checkmark & \checkmark & \checkmark & \checkmark & \checkmark & \checkmark & \checkmark & & \\
\bf Southampton, UK & & & & & & \checkmark & & & \checkmark\\
\bf Bristol, UK & \checkmark & & \checkmark & & \checkmark & & & & \\
\bf Glassgow, UK & & & & & \checkmark & \checkmark & & \checkmark & \checkmark \\
\bf Chicago, US & \checkmark & \checkmark & \checkmark & \checkmark & \checkmark & \checkmark & & \checkmark & \\
\bf Sacramento, US & \checkmark & \checkmark & \checkmark & \checkmark & \checkmark & \checkmark & & & \\
\bottomrule
\label{tab:smartcity}
\end{tabular}
\end{center}
\end{table*}
\vspace{-2mm}

\section{A reference IoT architecture for smart cities}\label{sec:refIoT}

In this section, we present one possible IoT architecture based on a $4$-layered stack that reflects modular design principles for smart city implementations.
\vspace{1mm}
\newline
\noindent
$\bullet$ \textbf{Layer-$1$: Physical/Virtual Space.}
The lowest layer is a collection of sensing elements, or data generators and consumers that provide \emph{context} information.
This information is sensed not only from the physical space of hardware-level sensors (that are deployed as part of static infrastructure or mobile entities), but also from soft sensors that exist in the virtual space such as aggregated event streams and crowd-sourced information.
\vspace{1mm}
\newline
\noindent
$\bullet$ \textbf{Layer-$2$: SNaaS.}
The SNaaS layer consists of control and data planes.
The data plane will offer channel models (e.g., event driven, sample and hold, etc.,) as a universal abstraction for input/output to the physical and virtual space of the sensing layer.
The control plane will be responsible for managing the sensor/ network and providing a standardized life-cycle for discovery, configuration and use of the channel models.
The combination is enabled by the creation of a TEDS-based plug-n-play infrastructure across platforms from multiple vendors.
This information will either be broadcasted by the new sensing entity to the infrastructure elements, or actively pulled by the infrastructure itself from a TEDS repository on receipt of the sensor identifier. 
Providing a self-describing mechanism to characterize the transducers will allow the SNaaS manager to record its properties and semantic description. 
The control plane will expose ways to turn on/off sensors or specific attributes, change their sampling interval, etc.,.
\newline
\indent
Following the configuration of the ``Thing'', the data plane will publish the observations of each sensor as individual data streams to the upper layer.
The combination of semantic capability and data push/pull (with a publish/subscribe model) will allow ``Things'' to be assembled into a semantically meaningful data flow graphs\cite{Zhouiswc:2012}.
The SNaaS layer will, therefore, act as a registry for semantic discovery and linkage of ``Things'', and for virtualizing the real world.
\vspace{1mm}
\newline
\noindent
$\bullet$ \textbf{Layer-$3$: ``Big-Little'' Data Management.}
SNaaS offers a conduit for streaming ``Little'' data from distributed physical and virtual sensors. 
This data has to be cataloged, curated; and if necessary, persisted and aggregated, in order to perform subsequent analytics. 
The data management layer helps discover and maintain a \emph{registry} of data sources and their characteristics such as periodicity, liveliness, quality, etc., and make them available for subsequent analysis. 
Access to data may be as streams of events pushed using a \emph{publish-subscribe} model, pulled on-demand, or accumulated at the device. 
These may be controlled using \emph{data privacy} and distributed access control policies that are enforced by this layer. 
\newline
\indent
Data management should also help integrate with aggregate and slow-changing ``Big'' data that is available from canonical sources or have accumulated from sensor streams. 
``Big'' data is hosted at central locations, globally or at local caches. 
Integrating ``Big'' with ``Little'' data requires a common vocabulary that helps to bridge the semantics.
Data quality checks and validation are needed to understand the usability of data for specific needs.
\newline
\indent
A \emph{data brokering service} that helps to interface data consumers with data producers is enabled using the building blocks of data ownership, incentives to collect data, data description, data quality and access control. 
The broker can help incentivize data collection and reuse through attribution, barter, or even monetary rewards.
\vspace{1mm}
\newline
\newline
\noindent
$\bullet$ \textbf{Layer-$4$: Analytics and Decision Making.}
Analytics for IoT applications often fall into categories of data and pattern mining, predictive analytics and forecasting, event and pattern detection, and optimizations. 
While there may be domain/application specific analytic, it is useful to have key analytics algorithms available that can be reconfigured to suit common needs. 
Time series and regression tree forecasting are useful for predicting future conditions.
Pattern mining and clustering can help group conditions that exhibit similar behavior so that collective action can take place, or extrapolation from one entity in the cluster to the rest can happen (e.g., recommendations)\cite{ddmm2011}.
Mining can also help identify causality between a pattern of features and an event of interest.
Such patterns and predictions can help feed into real-time complex event pattern matching\cite{govindarajanSJM14} that detect situations of interest and help
respond to, or preferable, preempt them. 
These also feed into optimization algorithms that can control physical or virtual ``Things'' to ensure reliability and efficiency of the system.
\newline
\indent
We refer our readers to \cite{IoT-IISc,IoT-India} for additional details about this reference IoT architecture. 

\begin{figure*}[p]
\begin{center}
\includegraphics[width=6.75in]{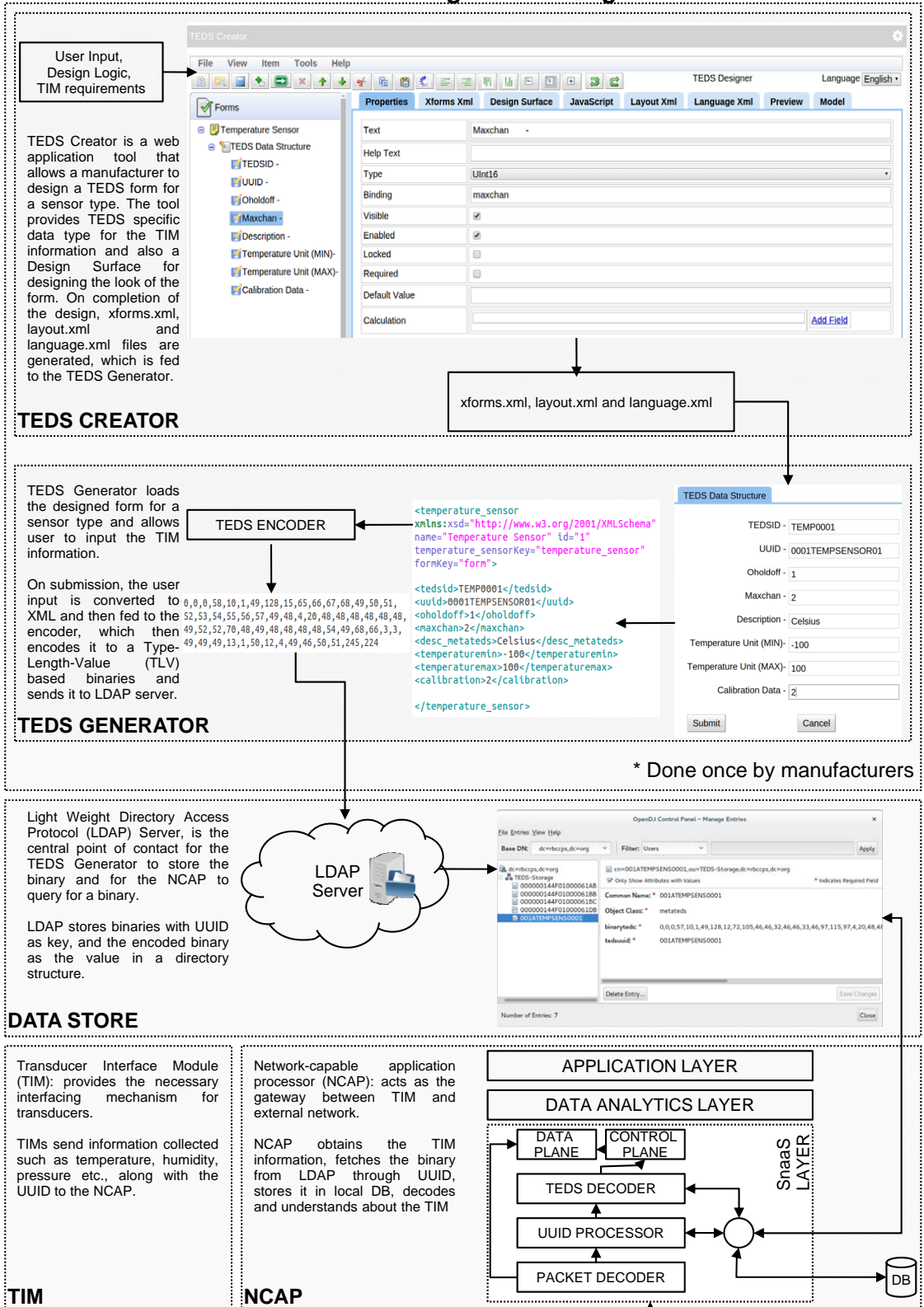}
\end{center}
\vspace{-5mm}
\caption{\textbf{SNaaS architecture and workflow.}}
\label{fig:snaas-architecture}
\end{figure*}

\section{The Design of SN\lowercase{aa}S}\label{sec:snaas}
In order to make this section self-contained, we first provide a brief overview of TEDS defined in the IEEE~$1451$ standard, and then present the overall architecture of SNaaS with preliminary microbenchmarks.

\subsection{Self describing devices with TEDS}
TEDS is the electronic version of the data sheet that we use to configure a sensor. 
TEDS brings forward the concept that if the data sheet is electronic and can be readily accessed upon sensor discovery, it would be possible
to configure the sensor automatically. 
This is analogous to the operation of plugging a mouse, keyboard, or monitor in the computer and using them without any kind of manual
configuration.
Thus, in principle, TEDS enables self-configuration of the system by self-identification and self-description of sensors and actuators (i.e., plug-and-play).
\newline
\indent
The IEEE $1451$\cite{1451} standard defines TEDS as a collection of multiple sections (such as sensor identification, calibration, correction data and manufacturer-related information) to form a generic electronic data sheet.
IEEE $1451$ divides the system into two general category of devices: (1) transducer interface module (TIM), which provide the necessary interfacing mechanism/functional units (such as the physical interface, signal conditioners, ADC/DAC convertors) for sensors and actuators; and (2) network capable application processor (NCAP), which acts as the gateway between the TIM and the external network.
TEDS can reside in TIM's memory, or at a central repository that is accessible by the NCAP from the network (referred to as Virtual TEDS).
The standard mandates the following four TEDS sections.
\vspace{1mm}
\newline
\noindent
$\bullet$ \textbf{Meta TEDS.} It contains referencing codes (for uniquely identification), different timing parameters (to detect the responsiveness of the TIM), and other information pertaining to the available transducer channels.
\vspace{1mm}
\newline
\noindent
$\bullet$ \textbf{TransducerChannel TEDS.} To enable proper functioning of the addressed transducer channel, various operational parameters (such as type, physical units of measurements, operating ranges, measure of various delays, etc.,) are specified in this TED.
\vspace{1mm}
\newline
\noindent
$\bullet$ \textbf{User's Transducer Name TEDS.} It contains the system referenced name (such as the model number and other manufacturer parameters) of the transducer.
\vspace{1mm}
\newline
\noindent
$\bullet$ \textbf{PHY TEDS.} It contains information about the physical connection between the TIM and the NCAP.
\vspace{1mm}
\newline
\noindent
The other TEDS section such as the calibration TEDS, Transfer Function TEDS, Frequency Response TEDs, etc., are optional; however, they can be used to increase the richness of the self-descriptive features of the transducer.
\newline
\indent
In the following section, we present the architecture of SNaaS, and use IEEE~$1451$ with certain design changes for IoT systems.

\subsection{System architecture}

The SNaaS plug-n-play system architecture consist of four components: TEDS Service, Lightweight Directory Access Protocol (LDAP), NCAP and TIM (Fig.~\ref{fig:snaas-architecture}).
\vspace{1mm}
\newline
\noindent
$\bullet$ \textbf{TEDS service and LDAP.}
The online TEDS service is a web application that is used to generate TEDS for TIMs, and comprises of two modules: ($1$) TEDS Creator (TC) and ($2$) TEDS Generator (TEDS-G).
The manufacturer is required to enter this information \emph{once} for registering the device.
\newline
\indent
Manufacturers can use the TC to generate a TEDS e-form template with fields related to the configuration parameters of the TIM. 
TEDS-G loads the e-form template, and allows the manufacturers to include details about the TIM.
On submission, an XML file with the TIM description is generated and is provided to the TEDS Encoder (TEDS-E), which encodes the TIM parameters into a Type-Length-Value (TLV) based binary (as per the IEEE $1451$ standard TEDS data block representation).
The encoded TEDS binary is stored in the LDAP directory service with the TIM UUID as the key.
\vspace{1mm}
\newline
\noindent
$\bullet$ \textbf{TIM and NCAP.}
TIMs with larger memory, communication bandwidth and continuous power source can store the respective TEDS in their memory, and it can be received by the NCAP on request. 
However, for resource constrained TIMs, the idea of virtual TEDS can be used for enabling self-describing capability. 
\newline
\indent
SNaaS, a service layer that offers a conduit for streaming data from distributed physical sensors, is implemented in the NCAP for plug-n-play functionality.  
NCAP listens on the specific medium of communication for TIM association.
The Packet Decoder (PD) parses the communicated information packet, following which the UUID processor checks with the local cache layer for TEDS history about the corresponding TIM.
If the information is not available locally, the LDAP server is queried with TEDS UUID as the key to obtain the binary TEDS followed by an update to the local cache. 
The decoded TEDS binary is passed to the control plane for performing the necessary transducer channel configurations.
Henceforth, the data plane provides the necessary abstraction for streaming the sensed information. 

\subsection{Microbenchmarks}

We have developed the SNaaS service layer for Android and Linux based NCAPs.
It was tested and evaluated with two types of TIMs: (a) custom designed CC$2540$\emph{Cheep} and (b) Arduino.
CC$2540$\emph{Cheep} uses Bluetooth Low Energy (BLE) as the communication primitive with the Android NCAP, while Arduino uses a USB Type~B wired communication mode with the Linux based NCAP.
Each of these TIMs were configured to send packets (containing the device UUID) every $100$\,msec.
\newline
\indent
Our preliminary microbenchmarks on latency measurements (for $50$~iterations) for the linux NCAP are presented in Table~\ref{tab:p1}. 
We considered two cases for this analysis.
\begin{itemize}
	\item Case~A: binary TEDS is \emph{available} in the local cache
	\item Case~B: binary TEDS is \emph{unavailable} in the local cache	
\end{itemize}
For Case~A, since the binary TEDS resides in the local cache, the total latency is only $4600$~$\mu$s.
For Case~B, the time taken for fetching the binary TEDS from the LDAP server, storing it in the local cache and decoding takes $900,000$~$\mu$s. 
It is to be noted that the LDAP is queried only \emph{once} for a particular TIM, after which the local cache of the NCAP is updated with the respective history. 

\begin{table}[t]
\begin{center}
\caption {Modulewise latency for Linux based NCAP} 
\label{tab:p1}
\normalsize
\begin{tabular}{lrr}
\toprule
\textbf{Module}  &  \textbf{Case~A ($\mu$s)}  &  \textbf{Case~B ($\mu$s)}\\
\midrule
Packet Decoder & $1300$ & $1300$ \\ 
UUID Processor & $2$ & $2$ \\ 
Local Cache & $3300$  & $1500$  \\ 
LDAP Query & $0$  & $897000$  \\ 
TEDS Decoder & $3$   & $3$  \\ 
\midrule
\textbf{Total} & \textbf{4600} & \textbf{900000} \\ 
\bottomrule
\end{tabular}
\end{center}
\vspace{5mm}
\end{table}

\section{Conclusion}
Interoperability is the key to manage systems of systems and to open markets to competitive solutions in IoT.
In this paper, we addressed the device-level interoperability problem prevalent in current IoT-based solutions for smart cities.
We proposed a SNaaS service layer, as part of a high level reference IoT architecture, that enables the creation of the plug-n-play infrastructure across platforms from multiple device vendors and implement it using the self-describing feature outlined in the existing IEEE~$1451$ standard.
Such a SNaaS layer enables transparent management of sensors and collection of observational data, and depending on the domain, the layer can be implemented using existing standards from IEEE or W$3$C.

\bibliographystyle{IEEEtran}
\bibliography{smartcities2015-references} 

\end{document}